\documentclass{jps-cp}
\usepackage{txfonts}
\usepackage{graphicx}

\title{Effects of Tensor Force in the Relativistic Scheme: A Case Study of Neutron Drops}

\author{Shihang \textsc{Shen},$^{1,2}$ Haozhao \textsc{Liang}$^*$,$^{2,3}$ Jie \textsc{Meng},$^{1,4,5}$ Peter \textsc{Ring},$^{1,6}$ and Shuangquan \textsc{Zhang}$^{1}$}

\inst{
$^{1}$State Key Laboratory of Nuclear Physics and Technology, School of Physics, Peking University, Beijing 100871, China\\
$^{2}$RIKEN Nishina Center, Wako 351-0198, Japan\\
$^{3}$Department of Physics, Graduate School of Science, The University of Tokyo, Tokyo 113-0033, Japan\\
$^{4}$Yukawa Institute for Theoretical Physics, Kyoto University, Kyoto 606-8502, Japan\\
$^{5}$Department of Physics, University of Stellenbosch, Stellenbosch, South Africa\\
$^{6}$Physik-Department der Technischen Universit\"at M\"unchen, D-85748 Garching, Germany}

\email{haozhao.liang@riken.jp}

\recdate{March 31, 2018}

\abst{
  Tensor force is an important component in the nucleon-nucleon interaction, nevertheless, the role of the tensor force in the spin properties in finite nuclei is much less clear.
  In this report, we mainly focus on our recent progress on this topic about the relativistic Brueckner-Hartree-Fock theory and the corresponding tensor effects on the spin-orbit splittings in neutron drops.
  This will serve as an important guide for future microscopic derivations of the relativistic and non-relativistic nuclear energy density functionals.
}

\kword{spin-orbit splitting, tensor effect, relativistic Brueckner-Hartree-Fock theory}

\begin{document}
\maketitle

\section{Introduction}

This proceeding and the presentation given in the conference are mainly based on our recent progress in the self-consistent relativistic Brueckner-Hartree-Fock (RBHF) theory for finite nuclei \cite{Shen2016, Shen2017}, the discovery of the tensor effects on the spin-orbit (SO) splittings in neutron drops \cite{Shen2018, Shen1802.08108}, and the discussion on the spin symmetry in the Dirac sea \cite{Shen1802.08110}.

To understand the nuclear energy density functionals in terms of nucleon-nucleon ($NN$) interaction is one of the present frontiers in nuclear physics.
As manifested by the quadrupole moment of deuteron, the tensor force is one of the crucial components in the $NN$ interaction \cite{Bethe1940}.
For heavier nuclei, the shell-model calculations showed that the tensor force plays an indispensable role in the shell-structure evolution in exotic nuclei far away from stability \cite{Otsuka2005}.
Also since around 10 years ago, intensive investigations on the impact of tensor forces have been carried out in the scheme of nuclear density functional theory (DFT), including both the non-relativistic and relativistic frameworks.
However, it is still difficult to pin down significant features in experimental data which are only connected to the tensor forces and therefore suitable for an adjustment of their parameters, as reviewed in Ref.~\cite{Sagawa2014}.

For example, it was found in Ref.~\cite{Long2007} that the two-proton separation energies of the $N = 82$ isotones can be reproduced only by the relativistic Hartree-Fock (RHF) theory with the effective interaction PKA1, which includes the meson-nucleon tensor interactions generated by both the $\pi$-pseudovector and $\rho$-tensor couplings, whereas such two-proton separation energies cannot be reproduced by the relativistic mean-field (RMF) theory, i.e., the relativistic Hartree theory, which excludes both tensor related couplings.
This implies fingerprints of tensor effects could be found in such observable related to the nuclear mass.
In contrast, it was found in Ref.~\cite{Lalazissis2009} that if one lets the $\pi N$ coupling strength as a free parameter to fit the nuclear masses on the nuclear chart globally, one obtains the best fit with vanishing tensor forces.

Furthermore, evolution of the single-particle spectra was the main motivation for most of the studies on the tensor effects during the past decade.
Famous examples are the energy differences between the $1h_{11/2}$ and $1g_{7/2}$ single-proton states along the $Z = 50$ isotopes as well as the energy differences between the $1i_{13/2}$ and $1h_{9/2}$ single-neutron states along the $N = 82$ isotones \cite{Schiffer2004}.
These empirical shell-structure evolutions could be described by the energy density functionals including the tensor force, e.g., by the non-relativistic Skyrme Hartree-Fock (SHF) theory with the effective interaction SLy5+T \cite{Colo2007}, and qualitatively by the RHF theory with PKO1 or PKO3 \cite{Long2008}.
The tensor forces were found to play important roles there, although in terms of tensor force the relativistic scheme uses slightly different language from that in the conventional non-relativistic scheme \cite{Tarpanov2008, Moreno-Torres2010}.
However, in the context of DFT, the single-particle energies are only defined as auxiliary quantities.
In experiment, they are often fragmented and therefore only indirectly accessible.
The fragmentation is caused by the effects beyond mean field, i.e., by the admixture of complicated configurations, such as couplings to the low-lying surface vibrations \cite{Litvinova2011}.

Another kind of observable deemed to be sensitive to the tensor force is the nuclear spin-isospin excitation.
For example, the spin-dipole resonances in $^{208}$Pb show a very special energy ordering among different angular-momentum components \cite{Wakasa2012}, i.e., $E_x(1^-) \lesssim E_x(2^-) < E_x(0^-)$, instead of the normal one $ E_x(2^-) < E_x(1^-) < E_x(0^-)$.
This could be described by the self-consistent SHF + random phase approximation (RPA) calculations with the effective interaction SLy5+T$_W$ or T43 \cite{Bai2010}.
Nevertheless, such particular spin-dipole resonances in $^{208}$Pb cannot be reproduced by any self-consistent relativistic RPA calculation so far, although the fully self-consistent RHF+RPA calculations reproduced nicely the spin-dipole resonances in another nucleus $^{16}$O \cite{Liang2008, Liang2012}.
This remains a puzzle for how to further improve the relativistic energy density functionals, in particular, the properties of their tensor component.

Because of the above aspects, obviously, how to determine precise values for the strength parameters of the tensor forces in universal nuclear energy density functionals by a phenomenological fit to experimental data in finite nuclei is still an open problem.
In such a situation we propose to determine these strengths from microscopic \emph{ab initio} calculations based on the well known bare $NN$ forces.

Very recently, we achieved for the first time the fully self-consistent RBHF theory for finite nuclei \cite{Shen2016, Shen2017}, and then we chose the systems of neutron drop to investigate and identify the essential tensor effects \cite{Shen2018}.
Since the proton-neutron interaction is not involved, the equations for neutron drops are much easier to be solved than those for finite nuclei.
Therefore they have been investigated in the literature by many different \emph{ab initio} methods \cite{Pudliner1996, Smerzi1997, Pederiva2004, Gandolfi2011, Bogner2011, Maris2013, Potter2014, Tews2016} and also by nuclear DFT \cite{Zhao2016}.

\section{Results and Discussion}

Starting from the relativistic bare $NN$ interaction Bonn A \cite{Machleidt1989}, the neutron drops confined in an external harmonic-oscillator potential are investigated by using the RBHF theory.
Special attention is paid on the possible signature of the tensor force in the neutron-neutron interaction, i.e., the evolution of the SO splitting with the neutron number.
See Ref.~\cite{Shen2017} for the detailed formalism and Ref.~\cite{Shen2018} for the numerical details.

\begin{figure}
\includegraphics[width=7cm]{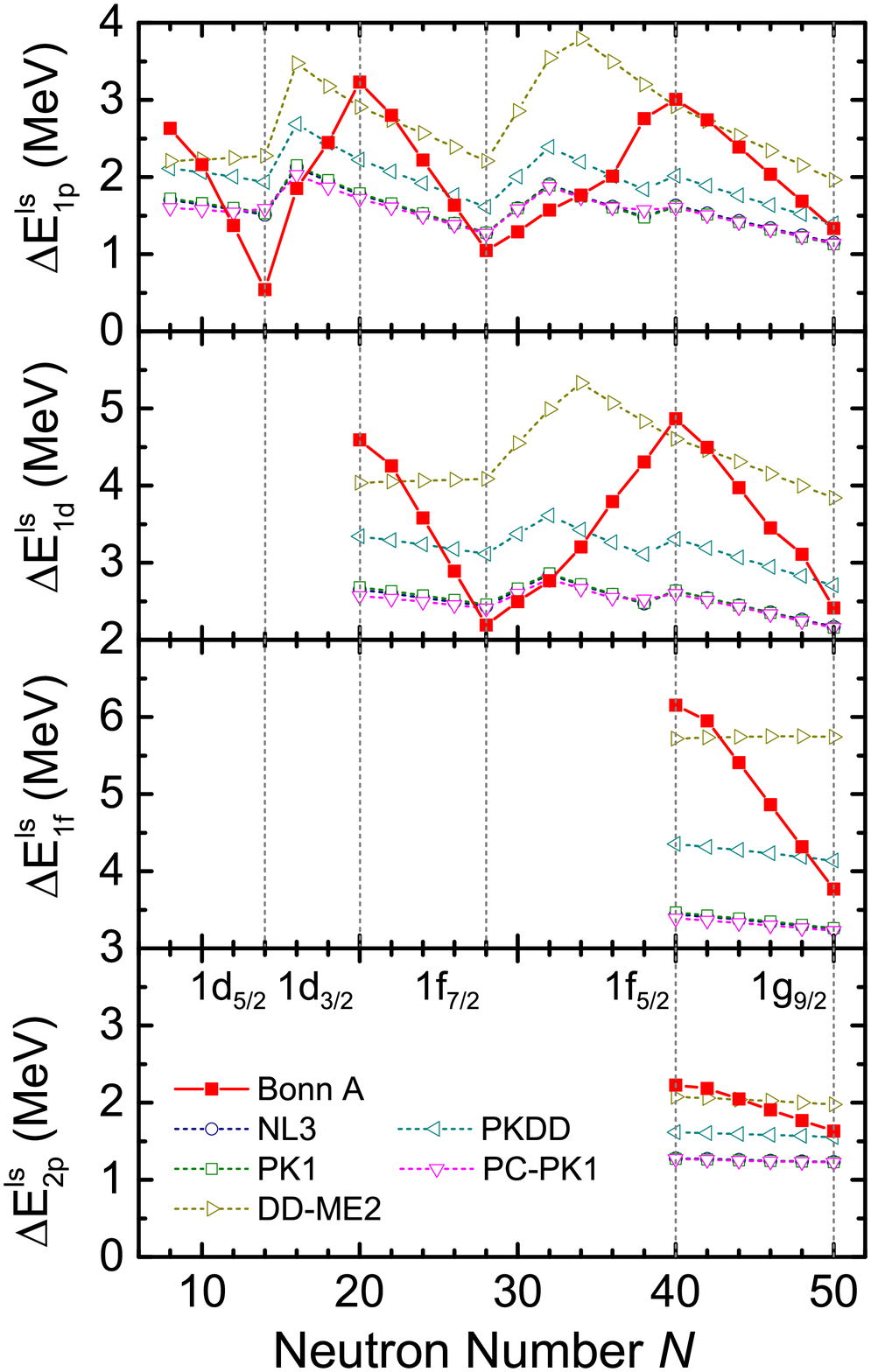}~~~~~~~~~~
\includegraphics[width=7cm]{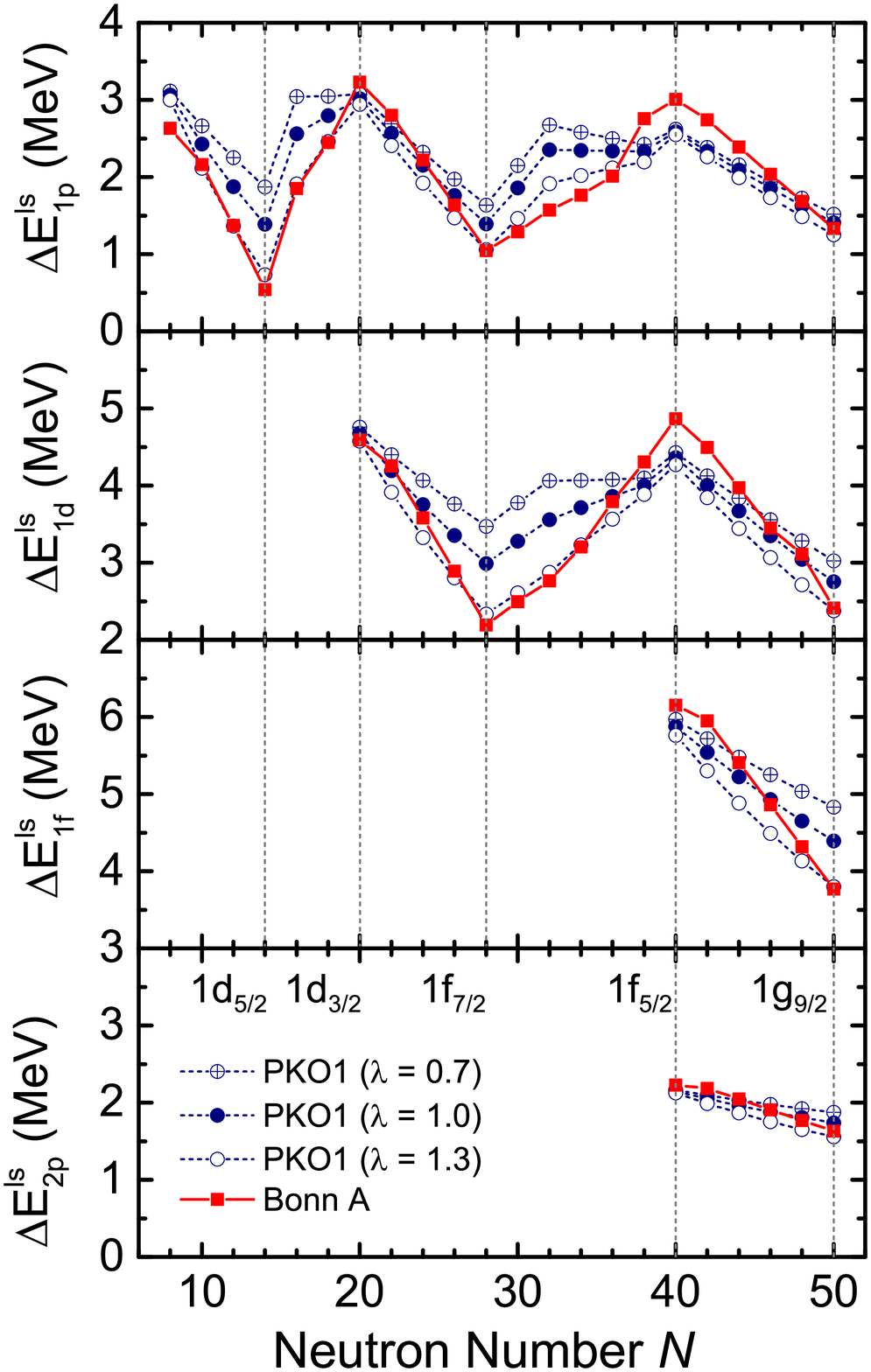}
\caption{(Color online) From top to bottom, spin-orbit splittings of the $1p,\,1d,\,1f$, and $2p$ doublets of $N$-neutron drops in a HO trap with $\hbar\omega = 10$ MeV, calculated by the RBHF theory with the relativistic bare nucleon-nucleon interaction Bonn A.
For comparison, (left) the results obtained by the RMF theory with the effective interactions NL3, PK1, PKDD, DD-ME2, and PC-PK1, and (right) the results obtained by the RHF theory with the effective interaction PKO1 with different $\pi N$ coupling strengths characterized by $\lambda$.
Taken from Ref.~\cite{Shen2018}.}
\label{fig1}
\end{figure}

First of all, we show in Fig.~\ref{fig1} the SO splittings of the $1p,\,1d,\,1f$, and $2p$ doublets of $N$-neutron drops in a HO trap calculated by the RBHF theory.
We can find a clear pattern from these \emph{ab initio} results.
Each SO splitting decreases as the SO aligned orbital with $j_>= l + 1/2$ is filled and reaches a minimum when this orbital is fully occupied, and then the SO splitting increases as the SO anti-aligned orbital with $j_< = l - 1/2$ begins to be occupied and reaches a maximum when this orbital is fully occupied.
This pattern repeats as the number of neutron continues to increase.

Otsuka \textit{et al.} \cite{Otsuka2005} have found a similar effect between neutrons and protons in nuclei.
They explained it in terms of the monopole effect of the tensor force, which produces an attractive interaction between a proton in a SO aligned orbital with $j_>=l + 1/2$ and a neutron in a SO anti-aligned orbital with $j'_<=l' - 1/2$, and a repulsive interaction between the same proton and a neutron in a SO aligned orbit with $j'_>=l'+ 1/2$.
They also pointed out that a similar mechanism, but with smaller amplitude, exists also for the tensor interaction between neutrons in the isovector $T=1$ channel \cite{Otsuka2005}.

That is exactly what is seen in the figure.
For instance, the SO splitting of the $1d$ doublets decreases from $N=20$ to $N=28$, because the interaction with the neutrons filling into the $1f_{7/2}$ shell above $N=20$ shifts the $1d_{5/2}$ orbital upward and the $1d_{3/2}$ orbital downward.
Above $N=28$, the neutrons filling into $2p_{1/2}$ and $1f_{5/2}$ interact with the $1d$-neutrons in the opposite way, and thus increase the SO splitting.
In other words, a clear signature of the tensor effect on the SO splitting in neutron drops is identified, for the first time, in the pure relativistic scheme.
Such a clear tensor effect can be served as a benchmark for the existing relativistic energy density functionals as well as as an important guide for their future developments.

In the left panel of Fig.~\ref{fig1}, we show the SO splittings of the same neutron drops calculated by various RMF energy density functionals, including the nonlinear meson-exchange models NL3 \cite{Lalazissis1997} and PK1 \cite{Long2004}, the density-dependent meson-exchange models PKDD \cite{Long2004} and DD-ME2 \cite{Lalazissis2005}, and the nonlinear point-coupling model PC-PK1 \cite{Zhao2010}.
It is seen that none of the results is able to reproduce the specific evolution pattern of the SO splitting generated by the \emph{ab initio} RBHF calculations, because it is known that there is no tensor component in the RMF energy density functionals.

In the right panel of Fig.~\ref{fig1}, we show the same results calculated by the RHF theory with the density-dependent meson-exchange model PKO1 \cite{Long2006}, which includes the tensor interaction generated by the $\pi$-pseudovector coupling.
It is seen that the specific evolution pattern of the SO splitting by the RBHF theory can now be clearly reproduced, although the size of the effect is somewhat too small.
This can be understood by the fact that in the original fit of PKO1 \cite{Long2006} only the information of nuclear mass was used, and thus the strength of the tensor force was basically out of control \cite{Lalazissis2009}.

In order to have more quantitative ideas on the proper strength of the tensor force, we multiplied a factor $\lambda$ in front of the $\pi N$ coupling, whereas other parameters in PKO1 remain unchanged.
It is seen that the size of the tensor effect does significantly depend on the value of $\lambda$, where the cases of $\lambda = 0.7,\, 1.0$, and $1.3$ are shown in the figure.
With $\lambda = 1.3$, the specific evolution pattern of the SO splitting generated by the \emph{ab initio} RBHF calculations can be nicely reproduced.
In such a way, the \emph{ab initio} calculations will guide the complete fit of the RHF energy density functionals in the near future.

\section{Summary and Perspectives}

We have studied the systems of neutron drop confined in an external HO potential using the RBHF theory with the relativistic bare $NN$ interaction.
It was found that the SO splitting decreases as the SO aligned orbital with $j_>= l + 1/2$ is being occupied, and increases again as the next SO anti-aligned orbital with $j_< = l - 1/2$ being occupied.
This is similar to the effects of the tensor forces between neutrons and protons found in Ref.~\cite{Otsuka2005}.
Such a specific pattern is not restricted to a specific mass region.
It is generally valid for all the neutron numbers $4 \leq N \leq 50$ under investigation.
Therefore, we can expect that a similar feature is valid also for realistic nuclei all over the nuclear chart.

This evolution pattern of the SO splitting can be reproduced by the RHF energy density functional with the tensor force, or even perfectly by readjusting a single parameter $\lambda$ for the tensor coupling strength, but cannot be reproduced by any RMF energy density functional.
In addition, it is important to remark that in the RBHF theory there are no higher-order configurations, which means the effects like particle-vibrational coupling \cite{Litvinova2011} are not included in these meta-data.
This allows to adjust the tensor force to these meta-data without the ambiguity of additional effects, because these effects are neither included in the present concept of nuclear DFT, nor in the present RBHF calculations.

Based on these recent progress, the ongoing research with different strategies include:
(1) to carry out a complete fit of relativistic energy density functional by taking into account the experimental data of nuclear masses and radii, as well as these \emph{ab initio} meta-data of the SO splittings in neutron drops to control the strength of the tensor force;
(2) to carry out a fit in each spin-isospin channel by using each spin-isospin density distribution generated by the \emph{ab initio} calculations, where the Kohn-Sham single-particle potentials and orbitals can be uniquely determined from the density distributions by the inverse Kohn-Sham method \cite{Wang1993};
(3) to derive the energy density functional directly from the bare $NN$ interactions in an \emph{ab initio} way \cite{Liang2018}.

\section*{Acknowledgements}

This work was partly supported by the Major State 973 Program of China No. 2013CB834400,
the NSFC under Grants No. 11335002, No. 11375015, and No. 11621131001,
the Overseas Distinguished Professor Project from Ministry of Education of China No. MS2010BJDX001,
the Research Fund for the Doctoral Program of Higher Education of China under Grant No. 20110001110087,
and the DFG (Germany) cluster of excellence ``Origin and Structure of the Universe'' (www.universe-cluster.de).
S.S. would like to thank the short-term Ph.D. student exchange program of Peking University and the RIKEN IPA project, and H.L. would like to thank the RIKEN iTHES project and iTHEMS program.

\end{document}